# Possible Manipulation of Kondo Effect by Transition between Antiferromagnetic and Ferromagnetic s-d Coupling

B. Q. Song, L. D. Pan, S. X. Du

*Institute of Physics, Chinese Academy of Sciences, P.O. Box 603, Beijing, 100190, China*

## Abstract

Kondo effect originates from antiferromagnetic (AFM) s-d coupling between magnetic impurity and the conduction electron, while it will be totally quenched in ferromagnetic (FM) regime due to malfunction of spin-flip. We investigate the possibility of switching on/off Kondo effect by transition of AFM/FM s-d coupling using 3d-metal phthalocyanine molecule (MPc) on Au(111) as a model system. A Hamiltonian model is constructed based on the feature of MPc molecule to show the condition for AFM/FM s-d coupling. The AFM s-d coupling could transform to FM s-d coupling if the spin state of the lowest unoccupied orbital changes

*Introduction.-* Introduction.-Probing and manipulating Kondo effect at the molecular/atomic level have attracted much investigation for application in spintronics [1-6]. With the technique of scanning tunneling microscopy/spectroscopy (STM/STS), diverse schemes of manipulation have been proposed in nano-systems, e.g. assemble of impurities [1], molecular conformations [3], attachment of ligands [6]. These schemes essentially follow a similar thinking: through a subtle modification of the system in a controllable way, notable changing of Kondo temperature $T_K$, would occur- this is seen as the key ingredient to realize switching devices. In present letter, we go beyond the $T_K$-centered manipulating scheme and propose manipulating Kondo effect by the transition of s-d exchange coupling from Antiferromagnetic (AFM) to Ferromagnetic (FM) type.

It is known that a series of novel behaviors in Kondo effect originate from AFM s-d coupling, which have been extensively investigated based on AFM s-d exchange model [8-13]. Whereas FM would lead to effective decoupling of local moment with conduction electron for low-energy behavior and thus quench Kondo effect [7]. Based on this fact, it is possible to switch Kondo effect on/off by tuning the s-d coupling type between AFM and FM in a well-chosen system. This mechanism is in principle distinct from $T_K$ manipulation, which merely tunes the coupling strength without changing s-d coupling type.

The s-d coupling arises from two key mechanisms: direct Coulomb coupling and virtual mixing. [14,15] Present work is confined to virtual mixing, which is normally dominant for transition metal (TM). Under this mechanism, AFM coupling $J_{k,k'}$ ($J_{k,k'} > 0$) can be derived from Anderson model [16] by a canonical transformation, namely, Schrieffer-Wolff transformation (SWT) [17]; whereas FM does not have a possibility to occur. In this letter, we make a subtle generalization of Anderson model, naturally including both possibilities of AFM and FM coupling. The generalization is based on features of a real system: 3d-TM Phthalocyanine (MPc) molecule adsorbed on Au(111) (Fig. 1a). With the model, we illustrate the origin of AFM/FM s-d coupling and present a scheme to make AFM/FM transition. Combining Density Functional Theory (DFT), we find that the AFM/FM transition can probably be realized for FePc on Au(111).

*Model*.-Features of isolated MPc molecules obtained from DFT [18] are summarized in Fig. 1b. Note that $d$-states of TM split due to the existence of ligand; in particular, $d_{x^2-y^2}$ turns to an empty orbital. Accordingly, we describe the general feature of three MPc molecules by a model $H$, which includes one single-occupied $d$-state $d_1$ with energy $E_1 (E_1 < 0)$ and one empty $d$-state $d_2$ with energy $E_2 (E_2 > 0)$. $d_1$, $d_2$ correspond to the topmost single occupied $d$-state and the empty $d$-state of TM respectively. For brevity, ligand orbitals do not explicitly appear in $H$, and their effects are incorporated into $E_1$, $E_2$. The gold substrate is described by a Fermi reservoir (FR), which couples with d-states through a mixing term. The total Hamiltonian yields: $H = H_0 + H_c + H_{ex} + V$

$$H_0 = \sum_{i,\sigma}^{i=1,2} E_i n_{i,\sigma} + \sum_{\mathbf{k},\sigma} \epsilon_\mathbf{k} c^\dagger_{\mathbf{k},\sigma} c_{\mathbf{k},\sigma} \tag{1}$$

$$H_c = \frac{1}{2} \sum_{i,\sigma}^{i=1,2} U n_{i,\sigma} n_{i,-\sigma} + U'\left(n_{i,\sigma} n_{i+1,\sigma} + n_{i,\sigma} n_{i+1,-\sigma}\right) \tag{2}$$

$$H_{ex} = -\frac{1}{2} J_0 \sum_{i,\sigma}^{i=1,2} n_{i,\sigma} n_{i+1,\sigma} \tag{3}$$

$$V = \sum_{\mathbf{k},i,\sigma}^{i=1,2} V_\mathbf{k} c^\dagger_{i,\sigma} c_{\mathbf{k},\sigma} + V_\mathbf{k}^* c^\dagger_{\mathbf{k},\sigma} c_{i,\sigma} \tag{4}$$

$c^\dagger_{\mathbf{k},\sigma}$, $c_{\mathbf{k},\sigma}$ create or annihilate an electron in FR with wave vector $\mathbf{k}$ and spin σ. $c^\dagger_{i,\sigma}$, $c_{i,\sigma}$ are corresponding operators for $d_i$ orbital. $n_{i,\sigma}$ is number operator. $U$ is Coulomb repulsion between the same $d$-state and $U'$ is that between different $d$-states ($U > U'$). $J_0$ is the exchange integral.

In the sum, we adopt the convention for indices $i$, 1+1=2, 2+1=1. The model formally resembles Anderson model but differs by including the empty $d_2$, which, we show shortly, has significant consequence for the s-d coupling type. Following Schrieffer's procedure [17], we commit a canonical transformation: $\widetilde{H} = e^{-S}He^{S}$ to eliminate the first order (is the solution of $V + [H_0, S] = 0$). Then expand the transformed Hamiltonian $\widetilde{H}$ on the $H_0$ bases. The matrix elements give,

$$\widetilde{H}_{ab} = E_a \delta_{ab} + \frac{1}{2}\sum_c \langle b|V|c\rangle\langle c|V|a\rangle \times [\frac{1}{E_a - E_c} + \frac{1}{E_b - E_c}] + h.c. \quad (5)$$

The sum of $|c\rangle$ includes all possible intermediate states. There are matrix elements, which correspond to spin-flip process: $\widetilde{H}_{+-} = \langle d_\uparrow|\widetilde{H}|d_\downarrow\rangle$, where $|d_\uparrow\rangle = c^\dagger_{k,\downarrow}c^\dagger_{1,\uparrow}|g\rangle$ and $|d_\downarrow\rangle = c^\dagger_{k,\uparrow}c^\dagger_{1,\downarrow}|g\rangle$. On the other hand, elements corresponding to non-spin-flip process are $\widetilde{H}_{++} = \langle d_\uparrow|\widetilde{H}|d_\uparrow\rangle$. It yields,

$$\widetilde{H}_{+-} = \sum_{\mathbf{k}} |V_\mathbf{k}|^2 \{\frac{1}{E_1 - \epsilon_\mathbf{k}} + \frac{1}{\epsilon_\mathbf{k} - E_1 - U}\} + h.c. \quad (6)$$

$$\widetilde{H}_{++} = \sum_{\mathbf{k}} |V_\mathbf{k}|^2 \{\frac{1}{E_1 - \epsilon_\mathbf{k}} + \frac{1}{\epsilon_\mathbf{k} - E_1 - U} + \frac{1}{\epsilon_\mathbf{k} - E_2^*}\} + h.c. \quad (7)$$

For convenience, we can set $\epsilon_\mathbf{k} = \epsilon_F = 0$ and define $E_2^* = E_2 + U' - J_0$ ($E_2^*$ has meaning of energy of $d_{2\uparrow}$ when $d_{1\uparrow}$ is occupied). As $V_\mathbf{k}$ are constant parameters, the magnitude of matrix elements is mainly determined by terms in the brace.

*Discuss two limiting cases.-* First, $E_2^*$ is a large positive value, i.e., $d_{2\uparrow}$ lies far above $\epsilon_F$. The effect of the extra $d_2$ becomes negligible and the model is reduced to Anderson model, which leads to AFM s-d coupling in the normal sense. Second, $E_2^*$ is a small positive value, i.e., $d_2$ lies close above $\epsilon_F$. In this limit, non-spin-flip element $|\widetilde{H}_{++}|$ becomes much larger than spin-flip element $|\widetilde{H}_{+-}|$ due to the extra term $1/(\epsilon_\mathbf{k} - E_2^*)$. This means that spin state of the impurity can hardly be changed by conduction electron, i.e. spin-spin interaction between impurity and substrate is effectively decoupled. Consequently, the impurity effect on conduction electron can be described by a spin-independent potential model, which leads to a virtual bounding state near Fermi level [7]. We employ FM to term this limit. It note that the FM exchange model is formally different from the pure potential model, since it contains spin exchange term. However, we use scaling procedure to show that FM exchange model is equivalent to a potential model in low-energy [22]. Thus, the exchange term, which leads to novel behaviors in AFM regime, is a merely trivial form in FM regime.

AFM and FM s-d coupling are the two limiting cases of the constructed Hamiltonian. Even it is impossible to draw a definite border between the two limits, we use $E_2^* > \max\{|E_1|, |E_1 + U|\}$ as the condition of AFM coupling for an intuitive picture. This algebraic condition actually corresponds to $d_{2\uparrow}$ lying higher than $d_{1\downarrow}$ (solid line in Fig. 1c). The condition for FM coupling is $E_2^* < \min\{|E_1|, |E_1 + U|\}$, i.e., $d_{2\uparrow}$ lying lower than $d_{1\downarrow}$ (Fig. 1d). Given that Anderson model consists of a single $d$-state, it is equivalent to putting $d_2$ infinitely high and unsurprisingly AFM occurs. To drive a system from one limit to the other, one has only to modify the relative arrangement of the two $d$-states. For instance, consider an initial AFM coupling and in some way $d_{1\uparrow}$ is moved toward $\epsilon_F$. Consequently $d_{1\downarrow}$ is also energetically raised given $U$ is constant; whereas $d_{2\uparrow}$ is less affected (for simplicity we neglect its shift in energy). When $d_{1\downarrow}$ overwhelms $d_{2\uparrow}$, AFM finally transforms to FM. (dashed line in Fig. 1c)

*Realistic calculation.-* We employ DFT to find a realistic scenario to make the AFM/FM transition. The first problem is to determine the s-d coupling type of a given system. A direct way is to evaluate the relative arrangement of $d_{1\downarrow}, d_{2\uparrow}$. However, DFT does not promise reliable levels above $\epsilon_F$. Thus, we turn to seek telltale signature for coupling type in ground state. The ground state shows that for all the three MPc (M=Mn, Fe, Co), The Au atom closest to TM is slightly spin-up polarized parallel to the moment of TM (Fig. 2a). The common feature implies spin-down electron effectively transferring from Au to the TM (Au has full d-shell and cannot adopt any more spin-up electron). Interestingly, due to receiving the transferred electron, the local moment increases for MnPc and FePc, whereas it decreases for CoPc (Table. 1). Increased local moment indicates spin-flip transfer: electron going from spin-down into spin-up states; whereas, decreased or unchanged local moment indicates non-spin-flip transfer. (Fig. 2b) We use a scattering process to interpret the transfer. Initially, electron fill the FR; when perturbation $V$ is turned on, electron in FR is scattered into $d$-state of TM. Then the transfer is connected to tunneling matrix $T_{\sigma,\sigma'} \equiv \langle d_\sigma | T(\omega) | \mathbf{k}_{\sigma'} \rangle$, the probability of electron tunneling from $\mathbf{k}$-state to $d$-state. The relative magnitudes of spin-flip matrix element $T_{+-} \equiv \langle d_\uparrow | T(\omega) | \mathbf{k}_\downarrow \rangle$ and non-spin-flip $T_{++} \equiv \langle d_\uparrow | T(\omega) | \mathbf{k}_\uparrow \rangle$ are significantly different in AFM and FM limits. We expand d-state in plane waves $|d_\sigma\rangle = \sum_\mathbf{k} A(\mathbf{k}) |\mathbf{k}_\sigma\rangle$.

$$T_{\sigma\sigma'} = \sum_{\mathbf{k}} A(\mathbf{k})^* \langle \mathbf{k}_\sigma | T(\omega) | \mathbf{k}'_{\sigma'} \rangle$$

$$T(\omega) = V + V G_0^+(\omega) V + V G_0^+(\omega) V G_0^+(\omega) V + \cdots$$

In AFM limit, it yields $|\langle \mathbf{k}_\uparrow | T(\omega) | \mathbf{k}'_\uparrow \rangle| \approx |\langle \mathbf{k}_\uparrow | T(\omega) | \mathbf{k}'_\downarrow \rangle|$ (particle-hole symmetry leads to equality). Then $T_{+-}$ should have a value comparable with $T_{++}$ ($A(\mathbf{k})^*$ is spin-independent and thus is identical for $T_{+-}$ and $T_{++}$). In FM limit, $|\langle \mathbf{k}_\uparrow | T(\omega) | \mathbf{k}'_\uparrow \rangle|$ primarily orders $1/(\epsilon_k - E_2^*)$. Then $|\langle \mathbf{k}_\uparrow | T(\omega) | \mathbf{k}'_\uparrow \rangle| \gg |\langle \mathbf{k}_\uparrow | T(\omega) | \mathbf{k}'_\downarrow \rangle|$, and thus $|T_{++}| \gg |T_{+-}|$.

The tendency (flip or non-flip) of transfer obtained from DFT just corresponds to the relative magnitudes of $T_{+-}$ and $T_{++}$. The spin-flip transfer occurring in MnPc and FePc implies that $|T_{+-}|$ makes notable contribution and thus should be of a value comparable with $|T_{++}|$; the non-spin-flip transfer for CoPc implies that $|T_{+-}|$ is much smaller than $|T_{++}|$. Hence, MnPc and FePc with a finite $|T_{+-}|$ correspond to AFM; CoPc with $|T_{++}| \gg |T_{+-}|$ corresponds to FM coupling. Note that $\mathbf{k}$- and $d$-states are sometimes taken as orthogonal for simper discussion [16]. Here, $A(\mathbf{k})$ means that there is finite probability of scattered plane waves entering $d$-state to contribute local moment. That is why the change of local moment connects the scattering tendency (flip or non-flip) and becomes an indicator for coupling type. It is convenient to denote $\Delta M$ as the local moment change before and after adsorption. $\Delta M > 0$ means increasing after adsorption and corresponds to AFM; $\Delta M < 0$ corresponds to FM.

The next problem is to realize the energy shift described in Fig. 1c. For this aim, a hydrogen atom is put right above the center of MnPc to bond with TM. H-bonding has two consequences. First, previously single-occupied $d_{z^2}$ of TM is paired up with the electron from H and consequently the local moment decreases by 1 $\mu_B$ (MnPc: 3 $\mu_B \to$ 2 $\mu_B$, FePc: 2 $\mu_B \to$ 1 $\mu_B$); in addition, $d_{z^2}$ becomes deep-lying (Fig. 3a,c) and can be neglected for present concern. Second, the unpaired $d$-states, e.g. $d_{xz}$ $d_{yz}$, undergo an energy shift toward $\epsilon_F$, which is about 0.5 eV for MnPc and 1 eV for FePc (Fig. 3b,d). H is chosen merely for simplicity and it can be replaced by other molecules, e.g. CO, NO. Bonding between H and TM is about 1.7 eV, stronger than molecule-substrate interaction (Table. 1), so it is reasonable to take H-bonding molecule, MnPc-H (2 $\mu_B$) FePc-H (1 $\mu_B$), as a whole to interact with the substrate. Note that $\Delta M$ is positive for FePc, whereas $\Delta M$ turns to negative for FePc-H. (Table. 1) The changing indicates that FePc undergoes a transition from AFM to FM after H-bonding. MnPc does not have such transition and it can be either due to less effective energy shift or the empty orbital lying too high.

*GGA+U.*-For present interest, Coulomb repulsion $U$ is the central parameter, since it can greatly influence orbital arrangement above $\epsilon_F$. Thus, we discuss $U$ in detail by performing GGA+U according to Dudarev's scheme [19], under which $U$ is an amendable parameter. A proper $U$ would correct the underestimated local moment in pure GGA. However, setting $U$ too large would push the system away from mixed valence regime and artificially leads to AFM coupling. $\Delta M$ versus $U$ is plotted in Fig. 4. For MnPc, FePc and H-MnPc, $\Delta M$ is insensitive to $U$. For CoPc, the local moment increases quickly when $U$ is added to 2 eV. This means that $\Delta M < 0$ in pure GGA is partially due to underestimation of Coulomb repulsion. However, when $U$ goes on increasing, $\Delta M$ increases little and remains in FM regime. This is interpreted as CoPc lies deep in FM regime, so a slightly oversize $U$ (reference $U$ for Co: 3 eV [20]) would not push it into AFM regime. Interesting things happen for FePc-H. At first, $\Delta M$ changes little, but when $U$=3 eV ($U$ for Fe: 1-2 eV [20]), $\Delta M$ drastically increases into AFM regime. Compared with CoPc, the distinct response of FePc-H to $U$ indicates that FePc-H lies in FM regime but near the border, thus is easily pushed into AFM regime by large $U$. The phase diagram of MPc family is given in Fig. 4.

*More discussion.*-Complexity of Kondo problem arises from the fact that full-range of length and energy scales contribute to physics. However, as long wavelengths locally make little difference in electron density and gradient, DFT (GGA) can hardly account the contribution from these length scales. Then a question arises as: whether DFT suffices to characterize Kondo features? The question is related to the fundamental assumption in this work: even DFT fails to reproduce Kondo states on the energy scale $k_B T_K$ and the length scale $v_F T_K/h$, it still provides reliable calculation on the energy scale $E_d$ and the length scale around atomic size, e.g. local charge displacement. The assumption is associated with universality [21], which has a significant consequence: effects on correlation scale (e.g. behavior near critical point) have little influence on atomic structure. The insensitive response of atomic scale makes it possible to estimate s-d coupling (mainly a local coupling) independently, as if effects on correlation length, e.g. Kondo screening states, do not exist. Then the local moment obtained from DFT can be interpreted approximately as the moment with screening cloud stripped.

Notes also that the use of ΔM to reflect the s-d coupling type is based on several simplifications. First, the interaction between molecule and substrate is taken simply as that between TM and its closest Au. This requires that molecular skeleton interacts weakly with the substrate. Second, we have interpreted spin polarization of Au as spin-down going off, instead of spin-up going into. This relies on full *d*-shell condition, which makes transfer analysis simple and independent of Wigner-Seitz radius. Thus using ΔM to reflect the s-d coupling type is most applicable to magnetic molecule on full d-shell substrate. Even ΔM reflects the s-d coupling only in qualitative sense, it has provides essential information about whether Kondo effect exists in a given system, allowing a direct comparison with experiments (sect 3 of [22]).

*Conclusion.*-We investigate the possibility of manipulating Kondo effect via AFM/FM transition of the s-d coupling. The coupling type depends on the relative arrangement of d-states and can be reflected by the change of local moment in weak interaction limit. Combining DFT, we find that AFM/FM transition can be realized in FePc by a H-bonding scheme. The scheme can be extended to CO-, NO-bonding, providing more chances to be assessed in experiment.

Figure

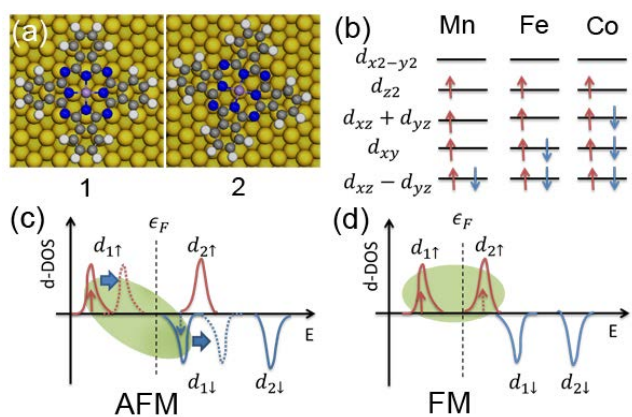

FIG. 1 (color online) (a) Configuration 1 and 2 of MPc molecule on Au(111). (b) The filling of *d*-states of isolated MPc molecules. The Three MPc molecules have similar orbital hybridization: $d_{z^2}$ $d_{xy}$ and $d_{x^2-y^2}$ remain as eigenstates; $d_{xy}$ and $d_{xy}$ form hybrid orbitals. Solid lines in (c) (d) are schematic arrangement of orbitals for AFM/FM coupling. Dashed line in (c) indicates a process of the single-occupied $d_{1\uparrow}$ and the empty $d_{1\downarrow}$ being energetically raised. When $d_{1\downarrow}$ overwhelms $d_{2\uparrow}$, AFM transits to FM.

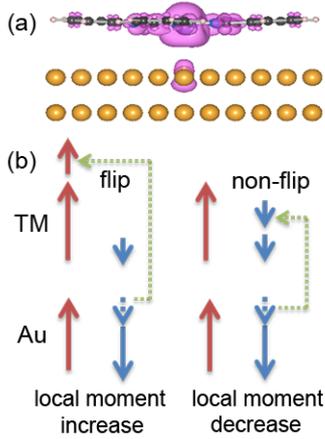

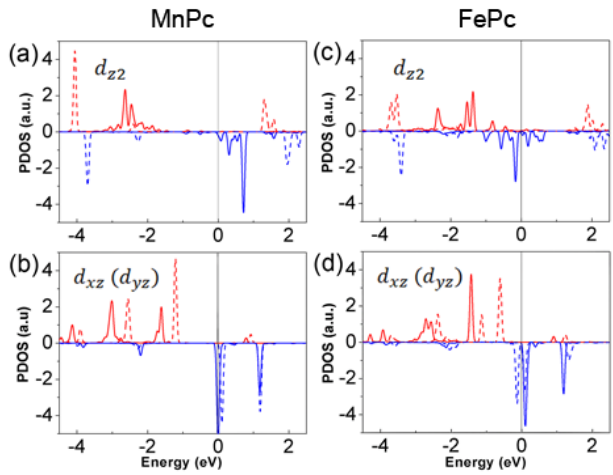

FIG. 2 (color online) (a) Spatial spin polarization ($\rho_\uparrow(\mathbf{r}) - \rho_\downarrow(\mathbf{r})$) of FePc on Au(111). The iso-surface is 0.0063 e/Å$^3$ (positive means spin-up dominant). Au atom closest to TM is spin polarized and the polarized moment is about 0.03 $\mu_B$. Spin polarization of other Au atoms is far smaller. (b) Schematic show for charge transfer between TM and Au. Increased local moment (left) indicates a spin-flip process: spin-down of Au atom to spin-up of TM. Decreased local moment (right) indicates a non-flip process.

FIG. 3 (color online) solid line in (a) (b) indicates on-site Projected-DOS of MnPc on Au(111) and dashed lines indicate that of MnPc-H on Au(111). (c) (d) are corresponding Projected-DOSs for FePc and FePc-H on Au(111). The shifts in (b) (d) are realistic correspondence to the process described in Fig. 1c.

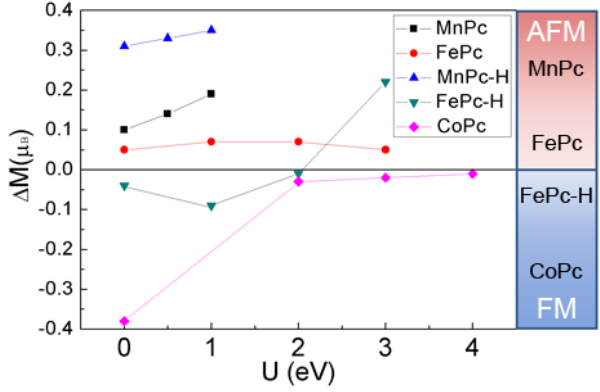

FIG. 4 (color online) ΔM indicates the local moment change of MPc before (isolated molecule) and after adsorption on Au(111). For comparison, we choose configuration 2 (more stable) for each molecule. U is chosen in vicinity of the reference value, (Mn: <1eV, Fe: 1-2eV Co: 3eV). At right panel, we give a schematic phase diagram of MPc family: MnPc CoPc lie deep in AFM and FM regime respectively; while FePc and FePc-H lies at the two sides of AFM/FM border.

| Mole. | Conf. | Adsorption E (eV) | moment (μB) | Couple type | Mole. | Conf. | Adsorption E (eV) | Moment (μB) | Couple type |
|---|---|---|---|---|---|---|---|---|---|
| MnPc | 1 | 0.465 | 3.35 | AFM | MnPc-H | 1 | 0.382 | 2.37 | AFM |
|  | 2 | 0.540 | 3.10 | AFM |  | 2 | 0.398 | 2.31 | AFM |
|  | isolated | -- | 3.00 | -- |  | Isolated | -- | 2.00 | -- |
| FePc | 1 | 0.374 | 2.01 | AFM | FePc-H | 1 | 0.324 | 0.99 | FM |
|  | 2 | 0.415 | 2.05 | AFM |  | 2 | 0.305 | 0.96 | FM |
|  | isolated | -- | 2.00 | -- |  | isolated | -- | 1.00 | -- |
| CoPc | 1 | 0.413 | 0.62 | FM | TPP-Co | 1 | 0.293 | 1.07 | AFM |
|  | 2 | 0.428 | 0.58 | FM |  | 2 | 0.305 | 1.21 | AFM |
|  | isolated | -- | 1.00 | -- |  | isolated | -- | 1.00 | -- |

TABLE 1

The adsorption energy, local moment and s-d coupling type for MPc on Au(111). The accuracy of DFT is about 10meV and U orders 1eV, thus the accuracy of local moment corresponds to 0.01. TBrPP-Co is reported to show Kondo effect on Au(111) [2]. We calculate it for comparison, showing AFM coupling in contrast to CoPc on Au(111) [5].

# Supplementary Materials

B. Q. Song, L. D. Pan, S. X. Du

*Institute of Physics, Chinese Academy of Sciences, Beijing, 100190, China*

## 1  The concept of s-d coupling

In this work, DFT calculation shows that for all the three MPc (M=Mn, Fe, Co), Au atom closest to TM is spin polarized parallel to the moment of transition metal (TM). It seems a paradox that the parallel spin alignment leads to AFM s-d coupling as we find in FePc, MnPc. The problem is related to the fact that s-d coupling is an effective spin-spin coupling, which cannot be directly interpreted in the sense of spin alignment between neighboring sites.

The concept of s-d coupling is closely related to two key mechanisms: direct coulomb coupling and virtual mixing. [1, 2] Direct coulomb coupling originates from exchange interaction between local *d*-state and conduction state. Since exchange interaction merely occurs between occupied states with parallel spin, direct coulomb coupling manifest itself as parallel spin alignment in the ground state. Virtual mixing, on the other hand, arises from quantum fluctuation: electron virtually hopping forth back between *d*-state and conduction states. Virtual mixing normally leads to AFM coupling, acting as a competing force against direct coulomb coupling. The coupling strength not only depends on the occupied level (ground state) also on empty orbital (intermediate states involved in the virtual process). The simplest calculation for virtual mixing is Schrieffer-Wolff transformation applied to non-degenerate Anderson model. The coupling strength yields, [3]

$$J_{k,k'} = V_k^* V_{k'} \{ \frac{1}{U + E_d - \epsilon_{k'}} + \frac{1}{\epsilon_k - E_d} \}$$

The magnitude of $J_{k,k'}$ depends on two denominators, which correspond to energy of occupied level and empty level respectively. When empty level is close to Fermi level, small denominator leads to large contribution from virtual mixing and AFM finally dominates. But no matter where the empty level locates, it doesn't qualitatively change the spin polarization or other properties in ground state. That means even in the case of virtual mixing being dominant, the local polarization can also be parallel aligned

The two mechanisms form the whole picture of s-d coupling. Spin polarization in the neighboring Au atom merely reflects one side of the picture: the FM contribution from direct coulomb coupling. Then it is becoming clear that the paradox at the beginning of this section is caused by equating direct coulomb coupling to s-d coupling; and is rationalized by including the virtual mixing mechanism. To find out the sign of the total coupling, we need precisely estimate coupling strengths under both mechanisms. However, when discussing AFM/FM transition, we have neglected the contribution of direct coulomb coupling and merely discuss it under virtual mixing. If we consider the correction from direct coulomb coupling, the AFM/FM transition would happen before $d_1$ overwhelm $d_2$ (Fig. 1c in the letter), making it easier to realize AFM/FM transition.

## 2 Calculation details of DFT

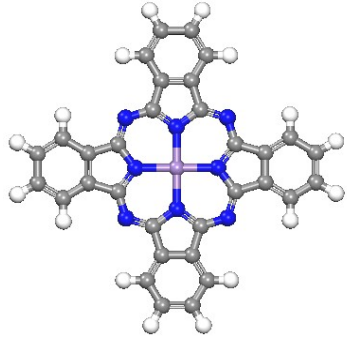

Figures 1 TM (Mn, Fe, Co) is in purple, hydrogen is in white, nitrogen is in blue and carbon is in grey.

The structure of MPc molecule is illustrated in the picture above. The constructed Hamiltonian model is based on the coexistence of single-occupied and empty orbital. In Fig. 1b of the letter, a schematic orbital occupancy picture is given with mere brief explanation. Here we present calculation details and the reason why we summarize a picture like that. We project wave function to onsite spherical Harmonics and get occupancy number of each $d$-orbital by integrating projected-DOS (PDOS) to Fermi level. The calculation shows that all $d$-orbitals are partially occupied. It is mainly caused by the central TM forming bonding with neighboring atoms. Rigorously, the eigenfunction is no longer merely the superposition of atomic $d$-states of the central metal atom, but mix with orbitals of neighboring atoms. As a result, $d$-state will be partially mixed into the anti-bonding state above Fermi level. Thus, the $d$-states become partially occupied.

The occupation number of each d-orbital

| Molecule | MnPc | | FePc | | CoPc | |
|---|---|---|---|---|---|---|
| Spin | up | Down | up | down | up | Down |
| $d_{x^2-y^2}$ | 0.43 | 0.32 | 0.43 | 0.36 | 0.46 | 0.41 |
| $d_{z^2}$ | 0.87 | 0.06 | 0.90 | 0.19 | 0.92 | 0.12 |
| $d_{xz}$ | 0.86 | 0.47 | 0.89 | 0.42 | 0.90 | 0.86 |
| $d_{xy}$ | 0.91 | 0.06 | 0.92 | 0.78 | 0.82 | 0.80 |
| $d_{yz}$ | 0.86 | 0.47 | 0.89 | 0.42 | 0.90 | 0.86 |

$d_{xz}$ and $d_{yz}$ coincide with each other, so the occupation numbers are identical.

The problem arises to us as how to draw the border between being occupied and empty. As for the orbital $d_{xy}$ of MnPc with 0.91 spin-up and 0.06 spin-down, the argument of the orbital being single-occupied will arise no doubt. But for $d_{x^2-y^2}$, the occupation number is 0.3~0.5, it deserves further explanation before taking the orbital as totally empty. From the viewpoint of physics, it is nonsense to artificially draw a line for occupation number, say 0.5, below which the orbital is considered empty. The argument is actually based on analysis of occupation number with varying TM species. From Mn to Co, the occupancy of $d$-state increases, but the occupancy of $d_{x^2-y^2}$ is almost unchanged. That suggests that main part of contributes to an anti-bonding state and the increasing $d$-electron would fill prior orbitals, like $d_{z^2}$ $d_{xz}$. Thus we expect $d_{x^2-y^2}$ is farther above the Fermi level and mention $d_{x^2-y^2}$ as empty. But more precisely, we should say that the eigenstate mainly contributed from $d_{x^2-y^2}$ is empty. For brevity, we mention the real eigenstate as $d_{x^2-y^2}$. The shortcut should not cause serious confusion in physical concept.

## 3  Comparison with empirical evidence

There has been some empirical evidence in support of conclusions obtained in this letter. In this sense, we shall stress the following experimental observation.

a)  CoPc on Au(111)

CoPc adsorbed on Au(111) has been studied by Zhao, et al. [4]. The essential conclusion is that no Kondo resonance is observed. The *dI/dV* spectra feature a broad peak near Fermi level, which doesn't broaden when temperature increases. Thus the author interprets it as the characteristic of $d_{z^2}$ orbital. In addition, they carry out a DFT (LSDA) study on the CoPc on Au(111). It shows that local moment of CoPc is totally quenched and thus the absence of Kondo resonance can be explained.

According to present work, CoPc on Au(111) corresponds to FM s-d coupling and the impurity would scatter conduction electrons quite like a pure potential. Thus, a virtual bonding state would form around Fermi level, which is temperature-independent and field-independent, just compatible with experimental observation. Essentially, we present another possible origin of the broad peak near Fermi level.

It should be addressed that even though the absence of Kondo effect for CoPc on Au(111) seems faithful, the theoretical explanation in [4] is worth reconsidering. We repeat the DFT study of CoPc also based on LSDA and the same model used in Zhao's paper. The key difference is surface Au atoms are fully relaxed in our study; however, all Au atoms of substrate are fixed by force in Zhao's work. We find that the artificial fixing leads to almost zero local moment, but full-relaxation leads to a local moment about 0.5 $\mu_B$. Considering there is no physical reason to fix surface Au atoms, our calculation seems more close to real situation. In addition, our GGA+U calculation shows that GGA+U correction is very important for CoPc on Au(111). It turns out that very small *U*~1eV would lead to a local moment up to 0.8 $\mu_B$. This indicates the quenching of local moment is probably due to underestimate U.

b)  TBrPP-Co on Au(111)

Interestingly, TBrPP-Co, which has similar structure as CoPc, is totally different from CoPc in Kondo effect. It is reported TBrPP-Co on Au(111) shows Kondo resonance state [5]. We carry out a DFT(GGA) study on TBrPP-Co as an comparison with CoPc. It turns out that TBrPP-Co corresponds to AFM coupling (Table 1 in the letter). The difference well explains the contrast observations of CoPc and TBrPP-Co on Au(111).

c)  FePc on Au(111)

Another experiment is FePc on Au(111) [6,7]. Even though there is controversy on Kondo temperature, the existence of Kondo resonance state seems of no problem. For our DFT study, FePc shows an AFM coupling type, also consistent with the experiment.

## 4 Coulomb-Hund term in Hamiltonian

The precise Coulomb & exchange interaction of the Hamiltonian in the literature should include following terms:

$$Coulomb: \frac{1}{2}U\sum_\sigma n_{1,\sigma}n_{1,-\sigma} + n_{2,\sigma}n_{2,-\sigma} + \frac{1}{2}U'\sum_\sigma n_{1,\sigma}n_{2,\sigma} + n_{2,\sigma}n_{1,\sigma} + n_{1,\sigma}n_{2,-\sigma} + n_{2,\sigma}n_{1,-\sigma}$$

$$exchange: -\frac{1}{2}J\sum_\sigma n_{1,\sigma}n_{2,\sigma} + n_{2,\sigma}n_{1,\sigma} \tag{1.1}$$

$U$ is the Coulomb interaction between the same $d$-orbital and $U'$ is that between different $d$-orbital. We adopt the convention for indices $i$, 1+1=2, 2+1=1, to put it in a more compact form

$$H_c = \frac{1}{2}U\sum_{i,\sigma}^{i=1,2} n_{i,\sigma}n_{i,-\sigma} + \frac{1}{2}U'\sum_{i,\sigma}^{i=1,2} n_{i,\sigma}n_{i+1,\sigma} + n_{i,\sigma}n_{i+1,-\sigma} \tag{1.2}$$

$$H_{ex} = -\frac{1}{2}J\sum_{i,\sigma}^{i=1,2} n_{i,\sigma}n_{i+1,\sigma} \tag{1.3}$$

Generally speaking, $U' < U$. If we make a little stronger assumption that $U' = 0$. Then to the second order, we obtain:

$$\widetilde{H}_{+-} = \sum_{\mathbf{k}}|V_{\mathbf{k}}|^2\{\frac{1}{E_1 - \epsilon_{\mathbf{k}}} + \frac{1}{\epsilon_{\mathbf{k}} - (E_1 + U)}\} + H.C. \tag{1.4}$$

$$\widetilde{H}_{++} = \sum_{\mathbf{k}}|V_{\mathbf{k}}|^2\{\frac{1}{E_1 - \epsilon_{\mathbf{k}}} + \frac{1}{\epsilon_{\mathbf{k}} - (E_1 + U)} + \frac{1}{\epsilon_{\mathbf{k}} - (E_2 - J)}\} + H.C. \tag{1.5}$$

Compared with the result of full interaction considered (Eq. 6,7), the difference is merely a constant shift of $E_2^*$. Since the third term of $\widetilde{H}_{++}$ always leads to a dominant contribution to the Hamiltonian element, the conclusion is qualitatively unaffected.

## 5 the relation between FM s-d exchange model and two-channel Anderson model

In this section, we would show that both FM s-d exchange model and two-channel Anderson model can be transformed into a pure potential form in the low energy limit.

(1) FM s-d exchange model

In this part, we show that FM s-d exchange model can be transformed to a pure potential model with renormalized parameters. The basic idea is to project out the impurity spin-down subspace and yield a form containing only one spin. Then estimate the influence of the spin-down subspace by estimating tunneling matrix elements that involve intermediate states in spin-down subspace. The effects of spin-down are absorbed in renormalized parameters.

The anisotropic s-d exchange model reads

$$H_{ex} = \sum_{\mathbf{k},\mathbf{k}'} J_h(S^+ c_{\mathbf{k},\downarrow}^\dagger c_{\mathbf{k}',\uparrow} + S^- c_{\mathbf{k},\uparrow}^\dagger c_{\mathbf{k}',\downarrow}) + J_z S_z(c_{\mathbf{k},\uparrow}^\dagger c_{\mathbf{k}',\uparrow} - c_{\mathbf{k},\downarrow}^\dagger c_{\mathbf{k}',\downarrow}) \tag{2.1}$$

Using scaling procedure to reduce the bandwidth $D$, we obtain the well-known scaling equation:

$$\frac{d(J_h)}{d(\ln D)} = -2\rho J_z J_h \ ; \quad \frac{d(J_z)}{d(\ln D)} = -2\rho J_h^2 \tag{2.2}$$

We learn from these equations: first, $J_z$ always increases with $D \to 0$; second, $J_h$ and $J_z$ obey a hyperbolic trajectory

$$J_z^2 - J_h^2 = A \tag{2.3}$$

$A$ is a constant characterizing the trajectory. Then there are three possible cases.

(i) $A > 0$. In this case, scaling drives $J_h$ to zero and $J_z$ to a finite renormalized value $\tilde{J}_z$, yielding an effective Hamiltonian

$$\tilde{H}_{ex} = \sum_{\mathbf{k},\mathbf{k}'} \tilde{J}_z S_z (c^\dagger_{\mathbf{k},\uparrow} c_{\mathbf{k}',\uparrow} - c^\dagger_{\mathbf{k},\downarrow} c_{\mathbf{k}',\downarrow}) \tag{2.4}$$

The Hamiltonian in the spin-up subspace can be expressed as follow,

$$H_{eff} = \sum_{\mathbf{k},\mathbf{k}'} V_z c^\dagger_{\mathbf{k},\uparrow} c_{\mathbf{k}',\uparrow} \; ; \quad V_z = \tilde{J}_z \langle s_z = \tfrac{1}{2} | S_z | s_z = \tfrac{1}{2} \rangle \tag{2.5}$$

By accounting the virtual processes that involve spin-down subspace, the coupling $V_z$ would be modified. The modification is related to

$$\langle d_\uparrow, \mathbf{k}_\uparrow | T(\omega) | d_\uparrow, \mathbf{k}'_\uparrow \rangle = \langle d_\uparrow, \mathbf{k}_\uparrow | (T^{(1)}(\omega) + T^{(2)}(\omega) + \cdots) | d_\uparrow, \mathbf{k}'_\uparrow \rangle \tag{2.6}$$

The $n$th-order of tunneling matrix reads

$$\langle d_\uparrow, \mathbf{k}_\uparrow | T^{(n)}(\omega) | d_\uparrow, \mathbf{k}'_\uparrow \rangle$$
$$= \langle d_\uparrow, \mathbf{k}_\uparrow | H_{eff} | c_1 \rangle \langle c_1 | H_{eff} | c_2 \rangle \cdots \langle c_{n-1} | H_{eff} | d_\uparrow, \mathbf{k}'_\uparrow \rangle \prod_{j=1,2\ldots n} \frac{1}{\omega - \epsilon_{c_j} + is} \tag{2.7}$$

The intermediate states $\{c_i\}$ should cover all possible virtual processes. The influence of the spin-down subspace corresponds to $\{c_i\}$ having at least one state in spin down subspace. In this case, it is evident that any spin down $c_j$ would cause either $\langle c_j | H_{eff} | c_{j+1} \rangle = 0$ or $\langle d_\uparrow, \mathbf{k}_\uparrow | H_{eff} | c_j \rangle = 0$. That means one spin subspace exerts no effects on the other and the two spins are thoroughly uncoupled. Thus, the Hamiltonian can be substituted by two replicas of pure potential models.

$$H_{eff} = \sum_{\mathbf{k},\mathbf{k}'} \tilde{V}_z c^\dagger_{\mathbf{k}} c_{\mathbf{k}'} \tag{2.8}$$

(ii) $A < 0$. The scaling drives $J_z$ to zero and $J_h$ to a finite value $\tilde{J}_h = \sqrt{-A}$, yielding an effective Hamiltonian, corresponding to (2.4)

$$\tilde{H}_{ex} = \sum_{\mathbf{k},\mathbf{k}'} \tilde{J}_h (S^+ c^\dagger_{\mathbf{k},\downarrow} c_{\mathbf{k}',\uparrow} + S^- c^\dagger_{\mathbf{k},\uparrow} c_{\mathbf{k}',\downarrow}) \tag{2.9}$$

In this case, the two spins are formally coupled. To decouple the two spins, we use the following linear transformation,

$$\begin{cases} c^\dagger_{k,\uparrow} = 1/\sqrt{2}(c^\dagger_{\mathbf{k},\leftarrow} + c^\dagger_{\mathbf{k},\rightarrow}) \\ c^\dagger_{\mathbf{k},\downarrow} = 1/\sqrt{2}(c^\dagger_{\mathbf{k},\leftarrow} - c^\dagger_{\mathbf{k},\rightarrow}) \end{cases} \begin{cases} c_{k,\uparrow} = 1/\sqrt{2}(c_{\mathbf{k},\leftarrow} + c_{\mathbf{k},\rightarrow}) \\ c_{k,\downarrow} = 1/\sqrt{2}(c_{\mathbf{k},\leftarrow} - c_{\mathbf{k},\rightarrow}) \end{cases} \tag{2.10}$$

$c^\dagger_{\mathbf{k},\leftarrow}, c^\dagger_{\mathbf{k},\rightarrow}, c_{\mathbf{k},\leftarrow}, c_{\mathbf{k},\leftarrow}$ are creation and annihilation operators with spin oriented in $x$ direction. Then, the Hamiltonian becomes

$$\tilde{H}_{ex} = \tfrac{1}{2} \sum_{\mathbf{k},\mathbf{k}'} \tilde{J}_h \{ (S^+ + S^-)(c^\dagger_{\mathbf{k},\leftarrow} c_{\mathbf{k}',\leftarrow} - c^\dagger_{\mathbf{k},\rightarrow} c_{\mathbf{k}',\rightarrow}) + (S^+ - S^-)(c^\dagger_{\mathbf{k},\leftarrow} c_{\mathbf{k}',\rightarrow} - c^\dagger_{\mathbf{k},\rightarrow} c_{\mathbf{k}',\leftarrow}) \} \tag{2.11}$$

In a similar fashion used in case (i), we write down the Hamiltonian confined in impurity spin-left subspace, yielding a Hamiltonian corresponding to (2.5)

$$H_{eff} = \frac{1}{2} \sum_{\mathbf{k},\mathbf{k}'} V_x c^\dagger_{\mathbf{k},\leftarrow} c_{\mathbf{k}',\leftarrow} \; ; \quad V_x = \tilde{J}_h \langle s_x = \tfrac{1}{2} | (S^+ + S^-) | s_x = \tfrac{1}{2} \rangle \tag{2.12}$$

Then, we examine the virtual processes using (2.6)(2.7). It is found that $V_x$ would not be influenced by the spin-right subspace. Thus, the spin-left and spin-right are thoroughly decoupled and the Hamiltonian can be written in a form like (2.8)

(iii) $A = 0$. This case is the rigorous isotropic situation. Scaling would lead to $\tilde{J}_z = 0$ and $\tilde{J}_h = 0$. That means impurity spin would thoroughly decouple from conduction electron. Thus, we can write Hamiltonian in a trivial potential form with $V = 0$.

Therefore, in the all three cases the s-d exchange term as the central part of AFM exchange model becomes a trivial form in FM regime, exhibiting equivalent behaviors to a pure potential model. It also notes that in case (ii) the scaling would drive $\tilde{J}_h$ to infinity. FM on the trajectory of $A < 0$ is not stable and would finally be driven to AFM regime. However, if the $|A|$ is small, the process is very slow, i.e., Kondo temperature is extremely low.

(2) two-orbital Anderson model

In this part, we examine the low-energy behavior of the two-channel Hamiltonian proposed in the present work. The Hamiltonian is given in Eq. (1)~(4) in the literature. In the similar fashion mentioned above, the effect of and the coupling is related to the tunneling matrix.

$$J_z: -\langle d_{1\downarrow}, \mathbf{k}_\uparrow | T | d_{1\downarrow}, \mathbf{k'}_\uparrow \rangle \sim S_z c^\dagger_{\mathbf{k},\uparrow} c_{\mathbf{k'},\uparrow} \ ; \ -\langle d_{1\uparrow}, \mathbf{k}_\downarrow | T | d_{1\uparrow}, \mathbf{k'}_\downarrow \rangle \sim S_z c^\dagger_{\mathbf{k},\downarrow} c_{\mathbf{k'},\downarrow} \qquad (3.1)$$

$$J_h: -\langle d_{1\uparrow}, \mathbf{k}_\downarrow | T | d_{1\downarrow}, \mathbf{k'}_\uparrow \rangle \sim S^- c^\dagger_{\mathbf{k},\uparrow} c_{\mathbf{k'},\downarrow} \ ; \ -\langle d_{1\downarrow}, \mathbf{k'}_\uparrow | T | d_{1\uparrow}, \mathbf{k}_\downarrow \rangle \sim S^+ c^\dagger_{\mathbf{k},\downarrow} c_{\mathbf{k'},\uparrow} \qquad (3.2)$$

Assuming $V_\mathbf{k}$ is insensitive to $\mathbf{k}$ and truncating to the second order of the tunneling matrix, we obtain

$$J_z = J_h = \frac{1}{N} \sum_\mathbf{q} \frac{|V|^2}{2} \left\{ \frac{1}{\epsilon_\mathbf{q} - E_1} + \frac{1}{\epsilon_\mathbf{q} + E_1 + U} \right\} \qquad (3.3)$$

Further assuming that $\epsilon_\mathbf{q}$ is very small compared with $|E_1|$ and $|U|$, we obtain the equivalent result of Schrieffer-Wolff transformation. When orbital $d_2$ is added, the tunneling matrix would be modified,

$$\langle d_{1\downarrow}, \mathbf{k}_\uparrow | T | d_{1\downarrow}, \mathbf{k'}_\uparrow \rangle = -\frac{|V|^2}{2} \sum_\mathbf{q} \left\{ \frac{1}{\epsilon_\mathbf{q} - E_1} + \frac{1}{\epsilon_\mathbf{q} + E_1 + U} + \frac{1}{\epsilon_\mathbf{q} + E_2^*} \right\} \qquad (3.4)$$

To examine the low-energy behavior, scaling is committed. $D < |\epsilon_\mathbf{q}| < D + \delta D$,

$$\delta J_z = -\langle d_{1\downarrow}, \mathbf{k}_\uparrow | T | d_{1\downarrow}, \mathbf{k'}_\uparrow \rangle = \frac{|V|^2 \rho_0 |\delta D|}{2} \left\{ \frac{1}{D - E_1} + \frac{1}{D + E_1 + U} + \frac{1}{D + E_2^*} \right\} \qquad (3.5)$$

When $E_2^*$ is very small compared with $|E_1|$ and $|U|$, the term $\frac{1}{D+E_2^*}$ would dominate the scaling ($D \to 0$) and drive $J_z$ much larger than $J_h$, yielding an equivalent situation as case (i) of FM s-d exchange model.